\documentclass[%
 reprint,
 amsmath,amssymb,
 aps,
]{revtex4-1}

\usepackage{graphicx}
\usepackage{dcolumn}
\usepackage{bm}
\usepackage{natbib}

\begin{document}


\title{Effects of geometric frustration in Kitaev chains}

\author{Alfonso Maiellaro}
\affiliation{%
	Dipartimento di Fisica ``E.R. Caianiello", Universit$\grave a$ di Salerno, Via Giovanni Paolo II, 132, I-84084 Fisciano
	(SA), Italy
}
\author{Francesco Romeo}%
\affiliation{%
 Dipartimento di Fisica ``E.R. Caianiello", Universit$\grave a$ di Salerno, Via Giovanni Paolo II, 132, I-84084 Fisciano
 (SA), Italy
}%
\author{Roberta Citro}%
\affiliation{%
	Dipartimento di Fisica ``E.R. Caianiello", Universit$\grave a$ di Salerno, Via Giovanni Paolo II, 132, I-84084 Fisciano
	(SA), Italy, \\CNR-SPIN, Via Giovanni Paolo II, 132, I-84084 Fisciano (SA), Italy
}%
\date{\today}

\begin{abstract}

We study the topological phase transitions of a Kitaev chain in the presence of geometric frustration caused by the addition of a single long-range hopping. The latter condition defines a legged-ring geometry (Kitaev tie) lacking of translational invariance. In order to study the topological properties of the system, we generalize the transfer matrix  approach through which the emergence of  Majorana modes is studied. We find that geometric frustration gives rise to a topological phase diagram in which non-trivial phases alternate with trivial ones at varying the range of the extra hopping and the chemical potential. Frustration effects are also studied in a translational invariant model consisting of multiple-ties. In the latter system,  the translational invariance permits to use the topological bulk invariant to determine the phase diagram and bulk-edge correspondence is recovered. It has been demonstrated that geometric frustration effects persist even when translational invariance is restored. These findings are relevant in studying the topological phases of looped ballistic conductors.
\end{abstract}

\pacs{Valid PACS appear here}
\maketitle


\section{\label{sec1}Introduction}

Recently, Majorana zero energy modes (MZMs) have attracted a lot of interest as they present promising properties for the implementation of fault-tolerant quantum computation\cite{doi:10.1146/annurev-conmatphys-030212-184337,alicea_2012,
Pachos2013}.
MZMs have been predicted to exist as zero energy states corresponding to localized edge modes in various condensed matter systems and some evidences have come from experiments conducted on semiconducting nanowires or ferromagnetic atomic chains proximized by a superconductor \cite{Mourik1003,Nadj-Perge602,PhysRevB.88.020407}. \\
The minimal model of a topological superconductor, where MZMs emerge at the edge of the system, is the Kitaev chain\cite{Kitaev_2001}, i.e. a model of spinless fermions subject to a $p$-wave superconducting pairing. After the seminal work of Kitaev\cite{Kitaev_2001}, various generalization of such model have appeared, either to describe coupled nanowires (Kitaev ladder)\cite{potter_2010,zhou_2011,nagaosa_2014,loss_2015,Maiellaro2018} or to include long-range pairing\cite{giuliano_2018} and disorder\cite{vonoppen_2011,romito_2011,dassarma_2013,altland_2013,wimmer_2014,PhysRevB.94.115166}. One of the major finding is that the topological phase is robust to disorder  but depends sensitively and non-monotonously on the Zeeman field, which is one of the ingredients required to stabilize a topological phase in nanowires with strong spin-orbit interaction.\\
Despite these numerous studies of the Kitaev chain, less is know about the effect of geometric frustration on the topological phase transitions. Although geometric frustration is being recognized as a new way of classifying magnets\cite{ramirez_nature}, its implementation in the topological context has not been discussed.  Thus here we  consider a Kitaev chain with an extra long-range hopping that realizes a legged-ring system (see Fig.1), the so-called Kitaev tie. In such a system a frustration between the state with two Majorana modes localized at the end of the legs and the state with hybridized modes along the tie emerges. The Kitaev tie model, whose geometry is depicted in Fig. 1, is not just a mere theoretical curiosity since it can be realized in looped single-walled carbon nanotubes\cite{doi:10.1021/nl0505997,doi:10.1021/nl901260b} where superconducting proximity effect can be easily implemented. For the legged-ring system the breakdown of translational invariance, induced by the extra hopping, does not allow to apply the bulk-edge correspondence and define a topological bulk index $Q$ \cite{PhysRevB.55.1142}. Thus, alternative approaches must be adopted.\\
Real space methods based on non-commutative geometry \cite{doi:10.1063/1.5026964} and on the wave function properties \cite{PhysRevB.94.115166,DeGottardi_2011} have appeared in the literature to characterize topological systems with broken translational invariance. In particular, the transfer matrix (TM) method, which is well known in  optics\cite{Zhan_2013}, has been widely used for 1D systems \cite{PhysRevB.29.1394,PhysRevB.61.9001} and is suited to reveal the emergence of localized Majorana zero energy states. It provides a complementary method to the calculation of the Pfaffian for systems with periodic boundary conditions and permits to deal with local disorder or impurities.\\
In this work, the topological phase transitions of the Kitaev tie are analyzed by using both the TM method \cite{PhysRevB.94.115166,DeGottardi_2011} generalized to the case of an extra long-range hopping, and  by calculating the Majorana polarization (MP) introduced in \cite{PhysRevLett.108.096802}. Both approaches provide a similar topological phase diagram, showing a rich interstitial-like behavior with non-trivial phases which alternate with the trivial ones when the chemical potential $\mu$ and the parameter $d$, controlling the hopping range, are varied. The geometric frustration strongly perturbs the energy spectrum of the system and the topological phase boundaries morphology reflects this perturbation. The interstitial-like character of the phase diagram arises as a result of the competition between the localizing effects at the edges of the chain and the hybridization of the Majorana modes along the ring, being this competition driven by interference effects. \\
In order to study topological frustration effects in translational invariant systems, a multiple-tie model is also investigated. In particular, when translational invariance symmetry is restored, the bulk-edge correspondence can be invoked to study the topological phase transitions by means of the bulk invariant, i.e. the Majorana number. \\
The paper is organized as follows: In Sec. \ref{sec2}, we introduce the Kitaev tie Hamiltonian, while its topological phase transitions are discussed in Sec. \ref{sec3}. In particular, the transfer matrix (TM) method, generalized to the presence of an extra long-range hopping,  is discussed in \ref{sec3.1}. The Majorana polarization is introduced in \ref{sec3.2} where the comparison between the topological phase diagram obtained by the TM method and the one obtained by the Majorana polarization is discussed. In Sec. \ref{sec4}, the multiple-tie system is investigated; there the topological phase diagram is obtained by using the Pfaffian invariant. Conclusions are given in Sec. \ref{sec5}. In the Appendices \ref{sec:hopping} and \ref{sec:example} the effect of the long-range hopping strength and the bulk-edge correspondence for a multiple-tie system are discussed.

\section{\label{sec2}The Kitaev tie model}
\begin{figure}
	\includegraphics[scale=0.55]{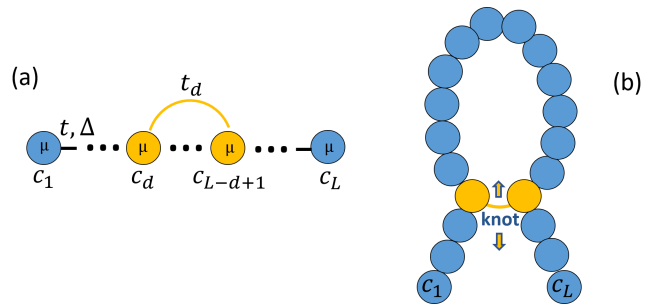}
	\caption{ (a) Tight-binding scheme of the Kitaev tie. (b) Toy-model representation.}
	\label{Model}
\end{figure}

A Kitaev tie is a Kitaev chain perturbed by the addition of a single long-range hopping linking two distant lattice sites. The tight-binding Hamiltonian of the model is:
\begin{equation}
H=H_K+H_d
\label{Hamiltonian}
\end{equation}
where $H_K$ is the usual Kitaev chain Hamiltonian \cite{Kitaev_2001}:
\begin{equation}
H_K=\sum_{j=1}^L[-\mu c^\dagger_{j}c_{j}+(\Delta c^\dagger_{j+1}c^\dagger_{j}-tc^\dagger_{j}c_{j+1}+h.c.)]
\label{Kitaev}
\end{equation}
written in terms of creation/annihilation  fermionic operators $c^\dagger_j/c_j$; $t$ and $\Delta$ are the hopping and the superconducting pairing amplitudes between nearest neighbor sites, $\mu>0$ is the chemical potential;  $H_d$ is the knot Hamiltonian linking the two sites $d$ and $L-d+1$:
\begin{equation}
H_d=-t_d( c^\dagger_{d}c_{L-d+1}+h.c.),
\label{KontHamiltonian}
\end{equation}
where $t_d$ is the hopping amplitude linking two distant sites. The range of the extra hopping, controlled by $d$, is varied to change the length of the legs (see Fig. \ref{Model}). A previous analysis of the Kitaev-tie energy spectrum in\cite{maiellaro2019topological} has already shown a frustration of the system emerging from a competition between localized edge modes and hybridized modes along the ring.
Moreover the breakdown of translational invariance symmetry leads to a system with no bulk associated\cite{doi:10.1142/S0129055X20300034}
since the long-range-hopping Hamiltonian $H_d=\sum_{k,q} c^\dagger_{k}V_{kq}c_q$, written in momentum representation, couples all $k$-modes via the single particle potential $V_{kq}=t_d\left[  e^{ikd}e^{-iq(L-d+1)}+e^{-iqd}e^{ik(L-d+1)}\right]$. Thus bulk-edge correspondence cannot be invoked and the topological phase transitions have to be analyzed by using real space methods. Accordingly,  in next section, we generalize the TM method introduced in Ref. \cite{PhysRevB.94.115166} to the legged-ring geometry.

\section{\label{sec3}Topological phase diagram of a Kitaev tie}
\subsection{\label{sec3.1}Transfer matrix approach with a long-range hopping}

\begin{figure}
\includegraphics[scale=0.46]{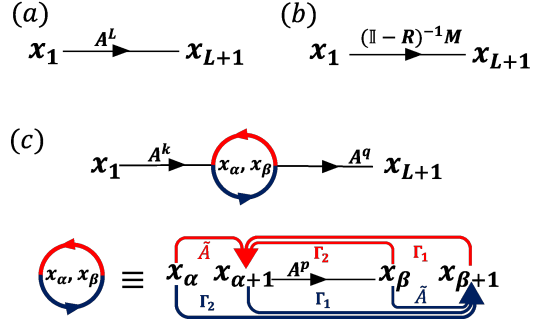}	
\caption{Sketch of the transfer matrix method (TM):(a) for a Kitaev chain and (b) for a Kitaev tie. (c) Operators loop structure of the TM in the presence of an extra hopping term connecting sites $\alpha$ and $\beta$. }
\label{TransferMatrixMethod}
\end{figure}

\begin{figure}
	\includegraphics[scale=0.33]{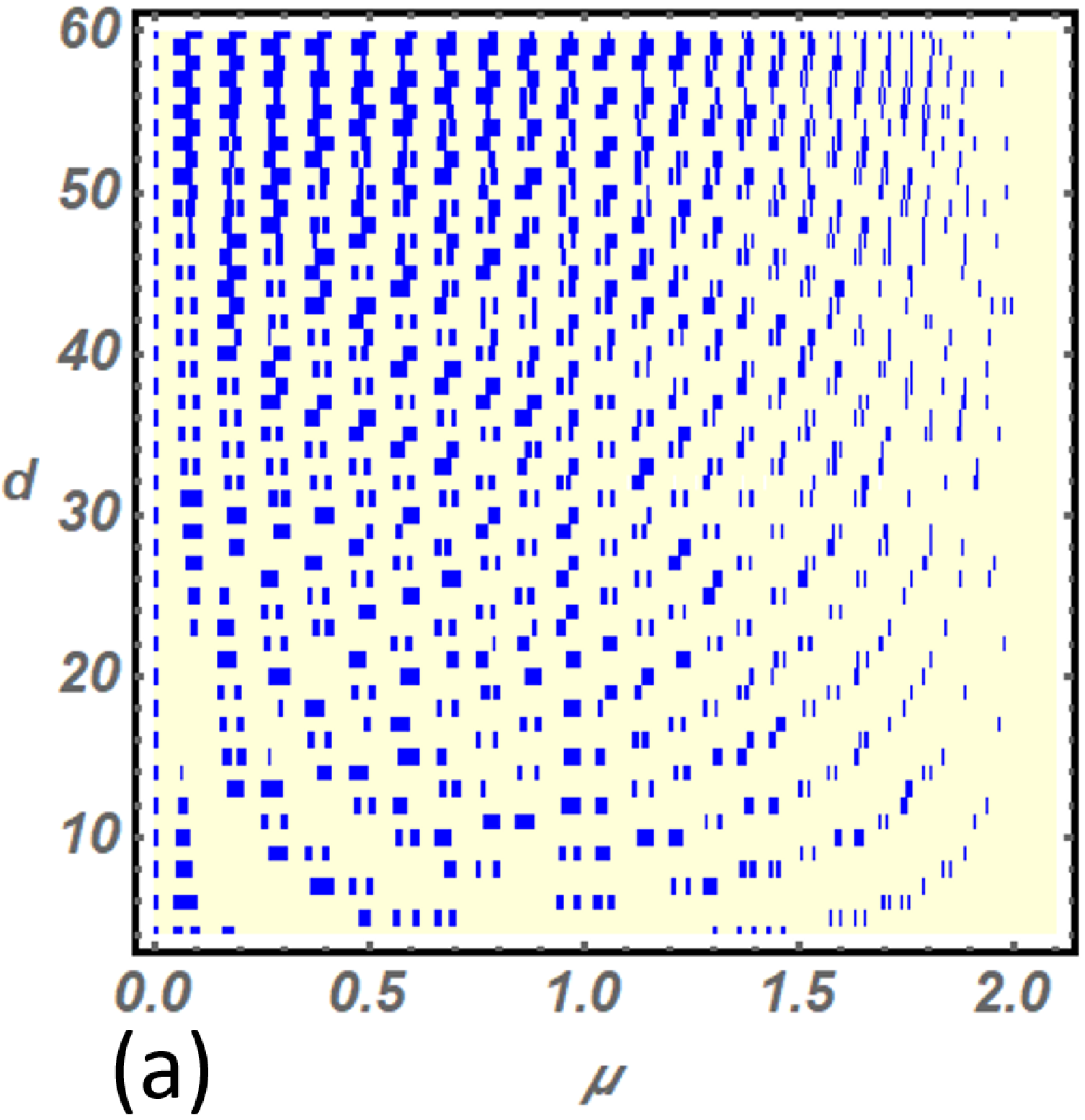}
	\includegraphics[scale=0.33]{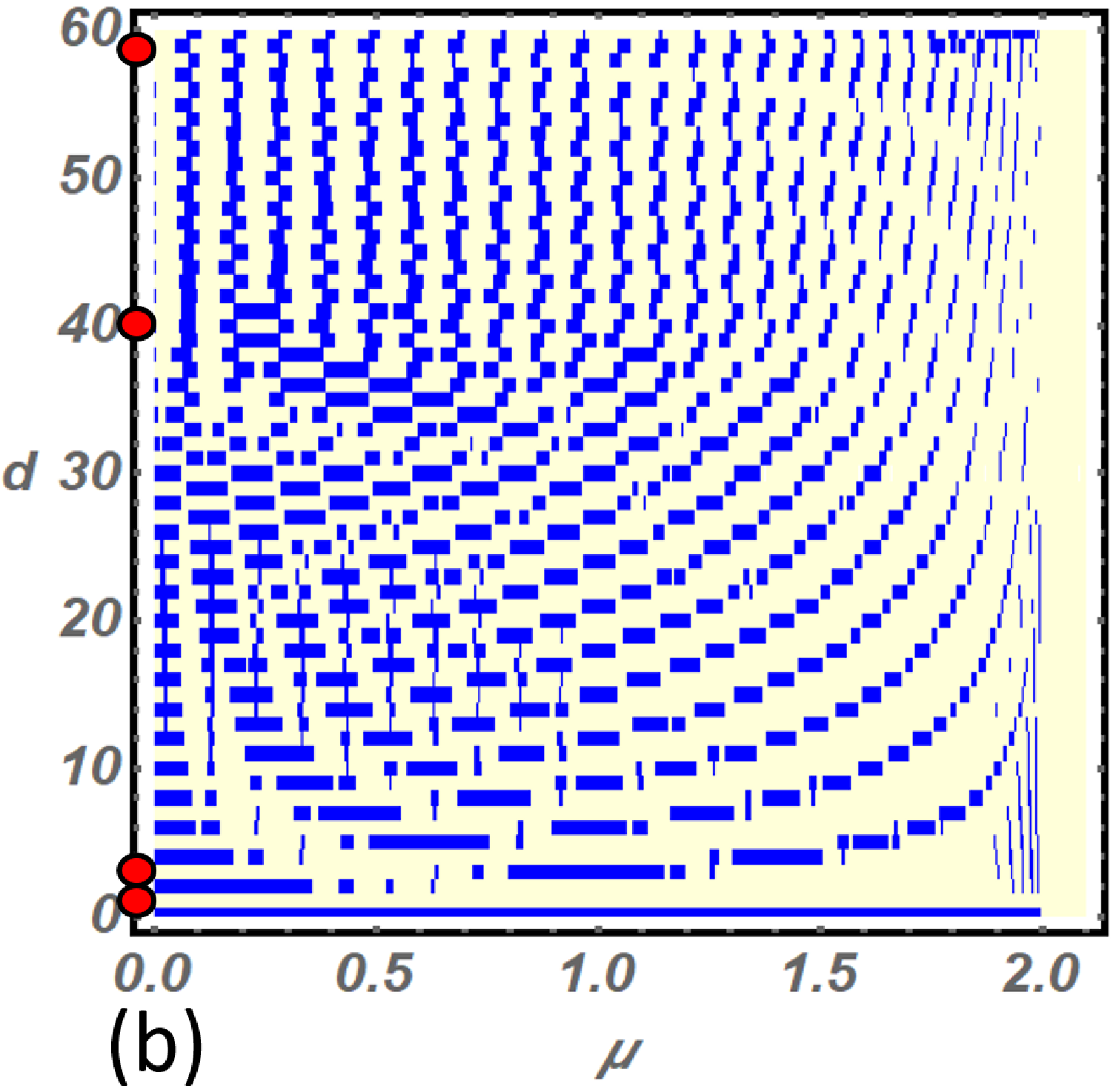}
\caption{Topological phase diagram of a Kitaev tie ($L = 121$) in the $d-\mu$ plane. The model parameters have been fixed as: $t_d=1$, $\Delta= 0.02$ in units of $t$. Panel (a) is obtained with the Majorana transfer matrix method ($T_{11}=10^{-7}$), while Majorana polarization has been used to obtain panel (b). The red dots of panel (b) correspond to four selected values of the extra hopping range controlled by the parameter $d$ ($d=1$, $3$, $40$, $59$) for which the polarization is plotted in Fig.4. Blue (white) regions represent topological (trivial) phases.}
\label{PD}
\end{figure}

Topological properties of finite-sized systems are usually described in terms of geometric indices also known as topological invariants $Q$, whose definition is strictly connected to the bulk of the system in which periodic boundary conditions are considered. The bulk-edge correspondence, then, can be invoked in order to calculate the number of zero energy edge modes \cite{doi:10.1142/S0129055X20300034}.
However, the topological properties of the frustrated system considered here cannot be addressed by means of the bulk-edge correspondence. Thus we base our analysis on the TM approach.\\
Starting from the Kitaev tie Hamiltonian, we make the change of basis from the fermionic operators $c_n$, $c_n^{\dagger}$ of Eq.(\ref{Hamiltonian}) to Majorana operators: $a_n=c_n+c^{\dagger}_n$, $b_n=i(c^{\dagger}_n-c_n)$ which satisfy the following  relations: $a^{\dagger}_n=a_n$, $b^{\dagger}_n=b_n$, $\{a_n,a_m\}=2\delta_{n,m}$, $\{b_n,b_m\}=2\delta_{n,m}$. In this new basis the Hamiltonian reads:
\begin{equation}
H_M=H_k^{'}+H_d^{'}
\label{MajoranaBasisH}
\end{equation}
where:
\begin{eqnarray}
H^{'}_k=&&-\frac{i}{2} \sum_{j=1}^{L-1}[(t-\Delta)a_jb_{j+1}-(t+\Delta)b_ja_{j+1}]\nonumber\\
&&-\frac{i}{2}\mu\sum_{j=1}^{L}a_jb_j\\
&&H^{'}_d=-\frac{i}{2}t_d(a_db_{L-d+1}+a_{L-d+1}b_d).\nonumber
\end{eqnarray}
The TM can be obtained by means of the Heisenberg equations of motion for the Majorana operators: $a_j(t)=a_j e^{-i\omega t}$ ($b_j(t)=b_j e^{-i\omega t}$) with $\hbar=1$: $\omega a_j=[a_j,H_M]$ ($\omega b_j=[b_j,H_M]$). Imposing the zero-energy constraint ($\omega=0$) for MZMs, two decoupled equations for the components of the Majorana wave functions $a_j$, $b_j$  are obtained ($a$- and $b$-mode equations):
\begin{eqnarray}
&&[a_j,H_M]= \nonumber\\
&&t_-b_{j+1}+t_+b_{j-1}+\mu b_j+t_d(\delta_{\alpha,j}b_{\beta}+\delta_{\beta,j}b_\alpha)=0\\
\label{b-modes}
&&[b_j,H_M]= \nonumber\\
&&t_-a_{j-1}+t_+a_{j+1}+\mu a_j+t_d(\delta_{\alpha,j}a_{\beta}+\delta_{\beta,j}a_\alpha)=0,
\label{a-modes}
\end{eqnarray}
where we have introduced the shortened  notations: $t_-=t-\Delta$, $t_+=t+\Delta$, $\alpha=d$, $\beta=L-d+1$.
The modes $a_j$ at different sites are related by the following equation involving the matrix $A$:
\begin{eqnarray}
\label{TMEq}
x_{j+1}=
A x_j+\delta_{\alpha,j}\left(
\begin{array}{cc}
-\frac{t_d}{t_+}a_\beta\\
0
\end{array}
\right)+\delta_{\beta,j}\left(
\begin{array}{cc}
-\frac{t_d}{t_+}a_\alpha\\
0
\end{array}
\right)\\
\nonumber
\end{eqnarray}
where:
\begin{eqnarray}
A=\left(
\begin{array}{cc}
	-\frac{\mu}{t_+}&-\frac{t_-}{t_+}\\
	1&0\\
\end{array}
\right),&&\ \ x_j=\left(
\begin{array}{cc}
a_j\\
a_{j-1}\\
\end{array}
\right).
\label{eq:modes}
\end{eqnarray}
In absence of the extra hopping term connecting the sites $\alpha$ and $\beta$, the model reduces to the standard Kitaev chain and the TM between the first and $L-$th site is simply the product of all the matrices $A$ between these two sites: $\mathcal{A}=A^{L}$ (see panel (a) of Fig. \ref{TransferMatrixMethod}).  The TM for  the $b$-mode has an identical structure with the change $t_-\rightarrow t_+$. Since for the Kitaev tie the sites $j=\alpha$ and $j=\beta$ are connected by the extra tunneling, the TM of the system has to take into account the more complex geometry and is not simply given by the product of A matrices.\\
Panel (c) of Fig. \ref{TransferMatrixMethod} shows the procedure to determine the TM of a Kitaev tie. In particular, the tie geometry introduces an operators loop structure in the TM equations which is reminiscent of the interference processes affecting the system response. To determine the TM of the Kitaev tie, first we specialize Eq. (\ref{TMEq}) to the cases $j=\alpha$ and $j=\beta$. Consequently, the following non-local relations are obtained:
\begin{eqnarray}
&&x_{\alpha+1}=\tilde{A} x_{\alpha}+\Gamma_{1} x_{\beta+1}+ \Gamma_{2} x_\beta\\
&&x_{\beta+1}=\tilde{A} x_{\beta}+\Gamma_{1} x_{\alpha+1}+ \Gamma_{2} x_\alpha
\end{eqnarray}
where the following auxiliary quantities have been introduced: $\tilde{A}=\left( \begin{array}{cc}\left( \frac{t_d^2}{\mu t_+}-\frac{\mu}{t_+}\right) &-\frac{t_-}{t_+}\\
1&0
\end{array}
\right)$, $\Gamma_1=\left( \begin{array}{cc}\frac{t_d}{\mu}&0\\
0&0
\end{array}
\right)$ and $\Gamma_2=\left( \begin{array}{cc}0&\frac{t_dt_-}{\mu t_+}\\
0&0
\end{array}
\right)$.
\\
Interestingly, the two terms $\Gamma_1$ and $\Gamma_2$ appear because of the long-range hopping.
On the other side, lattice sites $\alpha$ and $\beta$ are connected to first and last site of the chain by means of powers of the matrix $A$: $A^k$ and $A^q$, respectively (see panel (c) of Fig.\ref{TransferMatrixMethod}) thus we can rewrite the equations above as:
\begin{eqnarray}
\label{alphaMode}
&&x_{\alpha+1}=
(\mathbb{I}-\Gamma_2 A^p)^{-1} \tilde{A} A^k x_1+\nonumber\\ &&(\mathbb{I}-\Gamma_2 A^p)^{-1} \Gamma_1 x_{\beta+1},\\
\label{betaMode}
&&x_{\beta+1}= (\tilde{A}A^p+\Gamma_1) x_{\alpha+1}+\Gamma_2 A^k x_1,
\end{eqnarray}
where: $k=q=\alpha-1$, $p=L-2\alpha$. Finally, replacing Eq.(\ref{alphaMode}) into Eq.(\ref{betaMode}), the TM matrix  $T$ for the legged-ring model can be recognized:
\begin{eqnarray}
\label{Tfinal}
x_{L+1}=T x_1,
\end{eqnarray}
where $T=(\mathbb{I}-R)^{-1}M$ and the matrices $R$ and $M$ are given by:
\begin{eqnarray}
&&R=A^q (\tilde{A}A^p+\Gamma_1)(\mathbb{I}-\Gamma_2 A^p)^{-1} \Gamma_1 \left( A^{-1}\right)^q \nonumber\\
&&M=A^q(\tilde{A}A^p+\Gamma_1)(\mathbb{I}-\Gamma_2 A^p)^{-1}\tilde{A}A^k+A^q\Gamma_2A^k.
\nonumber
\end{eqnarray}
Once the TM is known, the topological phase transitions can be analyzed by imposing the localization requirement of the Majorana modes at the edge of the system, corresponding to the following condition:
$$a_{L+1}=T_{11}a_1+T_{12}a_0,\ \ {\text with}\ \ a_{L+1}=a_0=0,$$
or equivalently: $T_{11}=0$ \cite{note}.\\
The phase diagram obtained by the condition above is shown in panel (a) of Fig. \ref{PD} for a tie of $121$ sites at varying the chemical potential $\mu$ and the extra hopping range, controlled by $d$. Topological phases (blue regions) nucleate inside trivial regions (white regions). Moreover, the number of non-trivial phases increases when the circumference of the ring is reduced ($d$ is increased) i.e. when the system approaches a perturbed Kitaev chain limit and $\beta=\alpha+2$. The interstitial character of the topological phase is more evident when $d$ is lower than a critical value since the system is similar to a ring with very short legs.

\begin{figure}
	\includegraphics[scale=0.3]{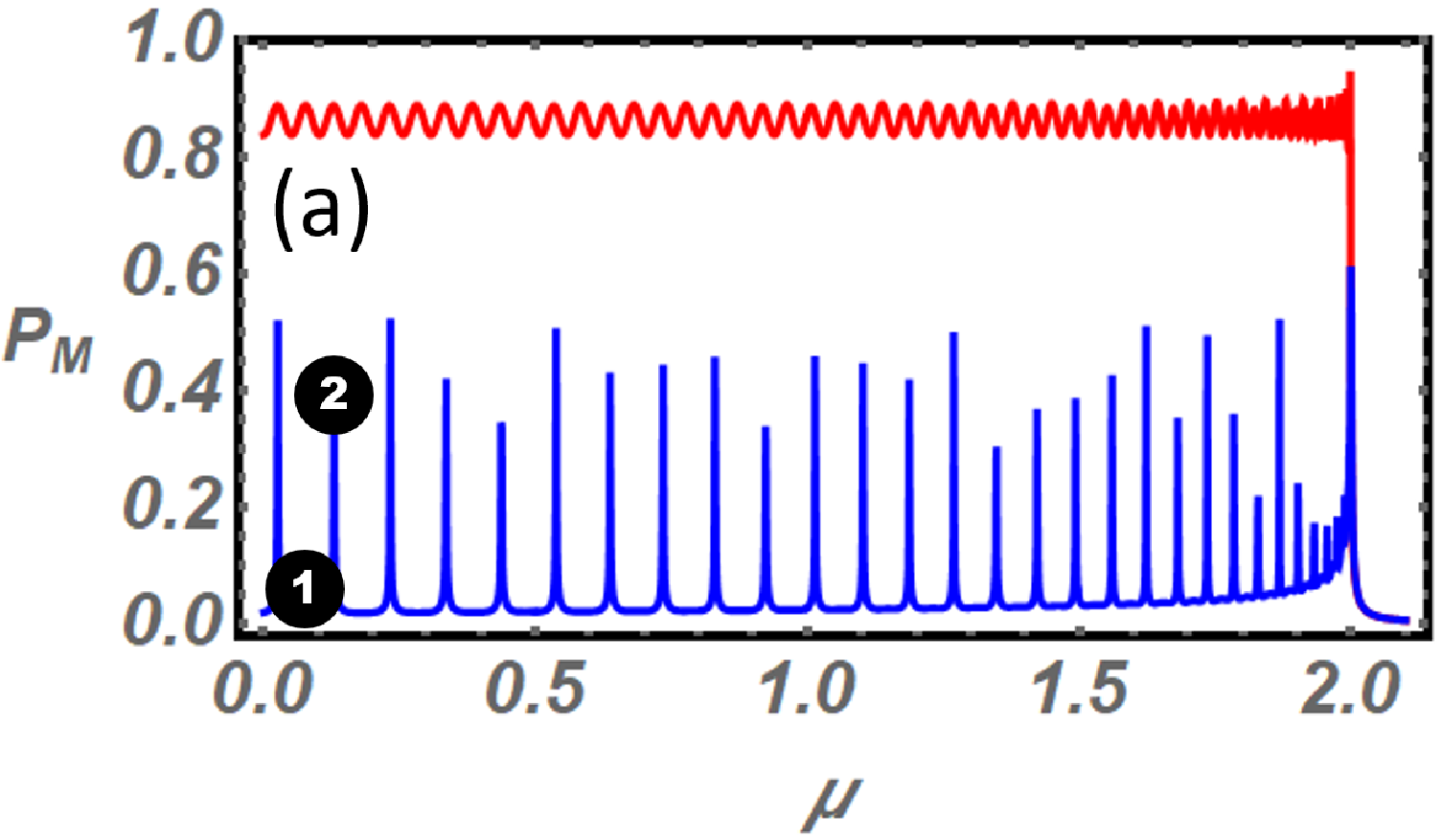}
	\includegraphics[scale=0.3]{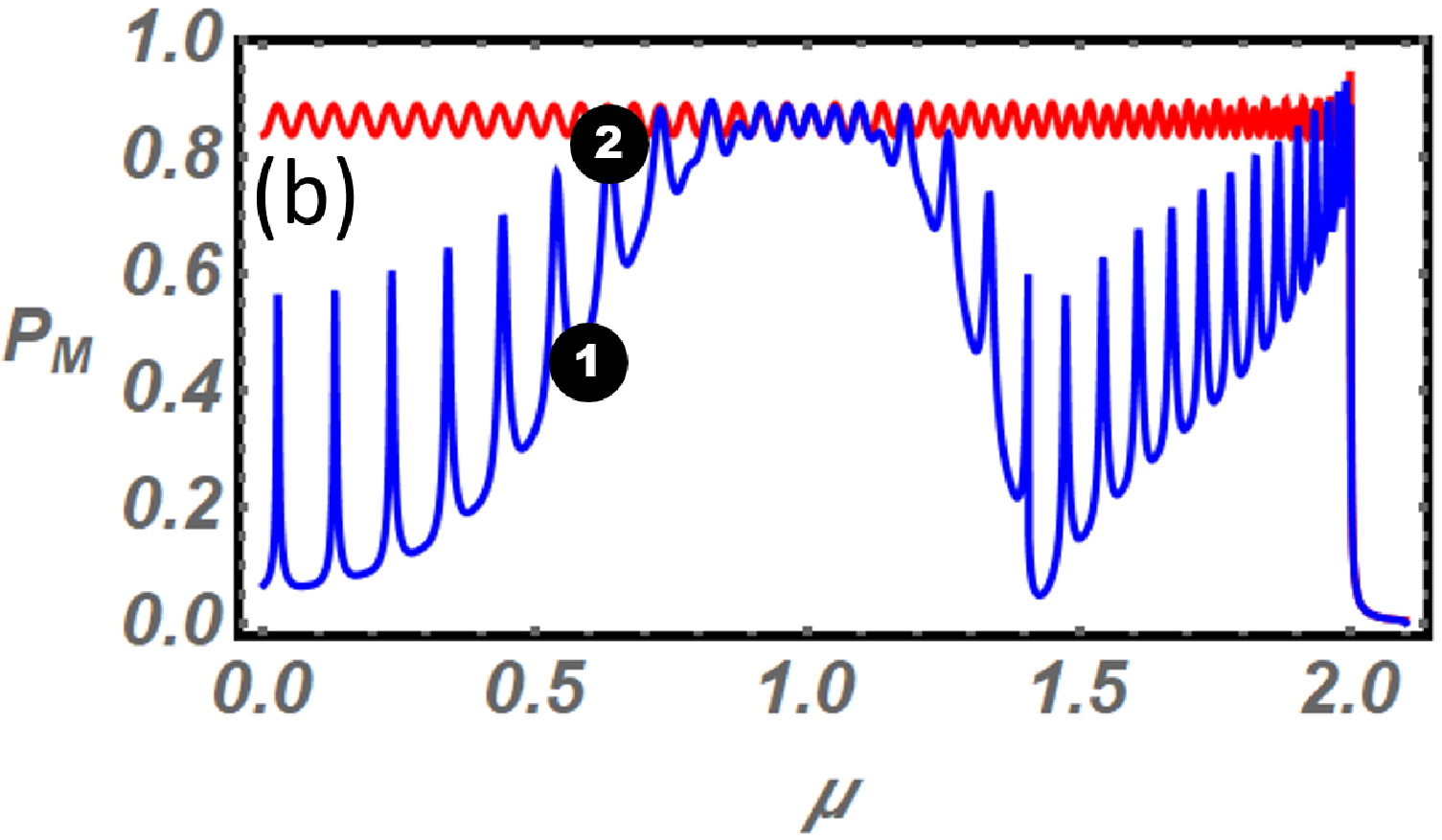}\\
	\includegraphics[scale=0.3]{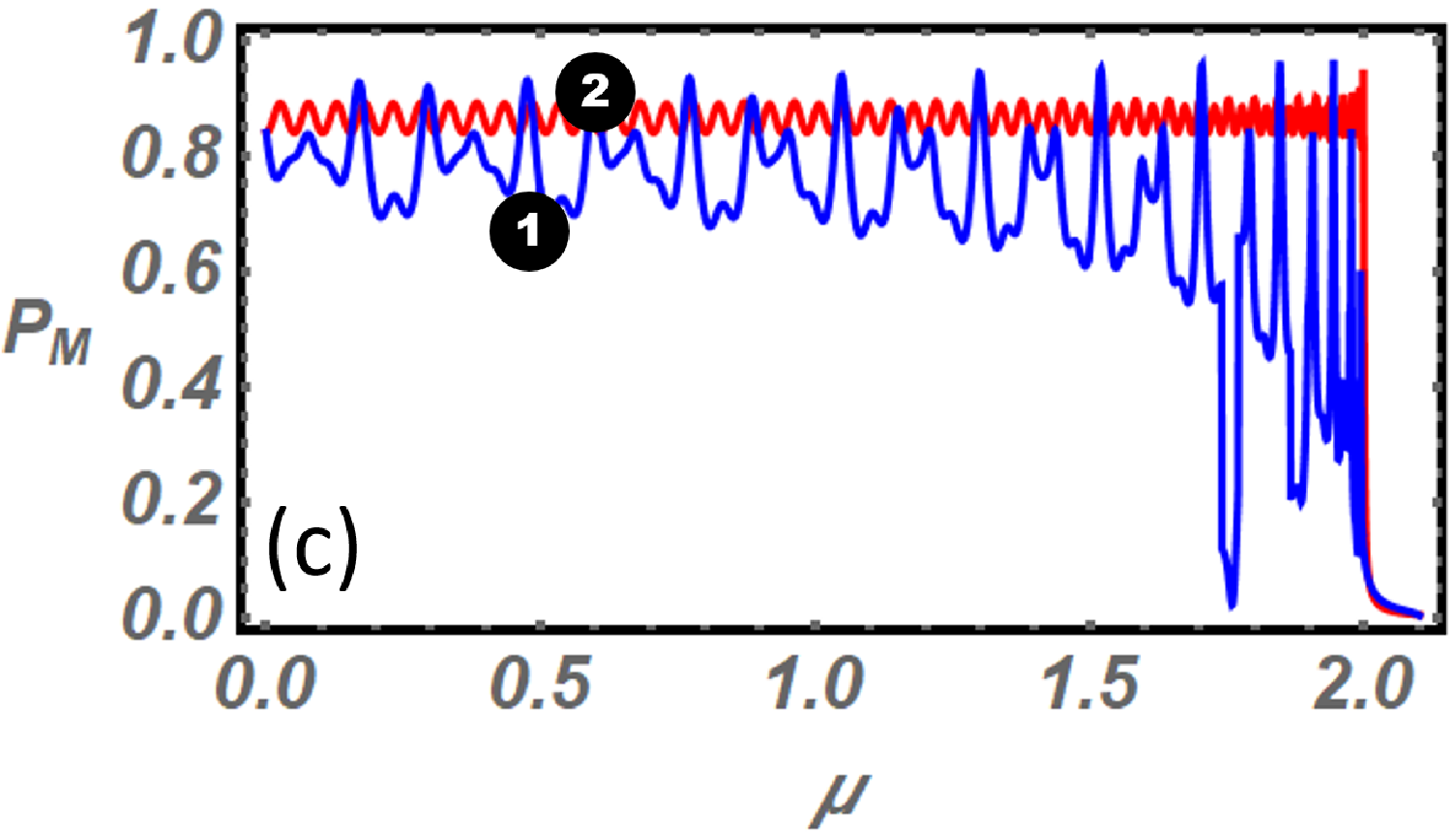}
	\includegraphics[scale=0.3]{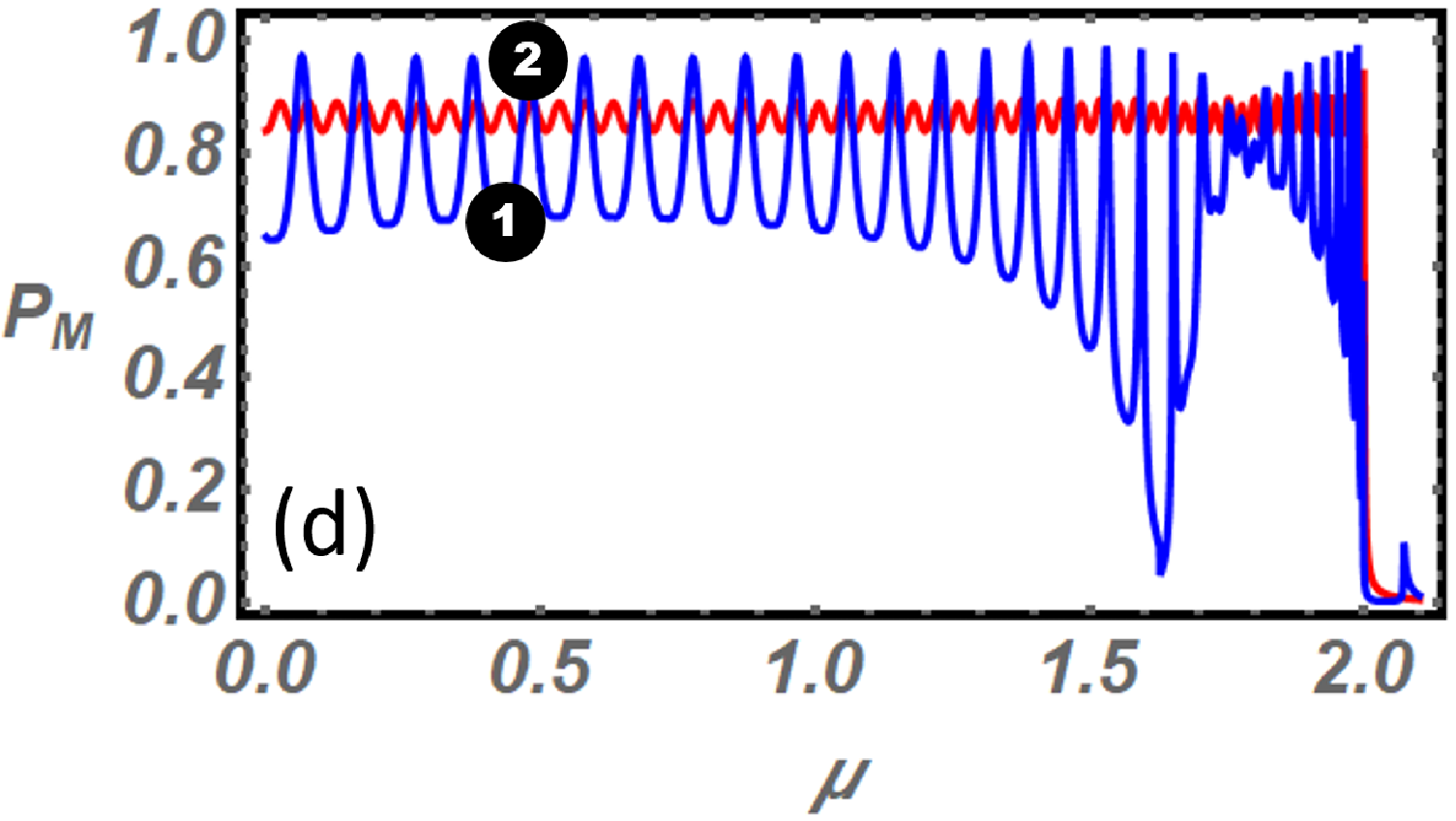}
	\caption{Majorana polarization (MP) of the Kitaev tie (blue curves) for the values of $d$ corresponding to the red dots of Fig. \ref{PD}. The cases of a Kitaev ring ($d=1$, panel (a)), a quasi-Kitaev ring ($d=3$, panel (b)), $d=40$, panel (c) and of a perturbed Kitaev chain ($d=59$, panel (d)) are shown. The red curves in each panel represent the MP of a Kitaev chain of the same size $L=121$. The numbered black circles are the selected minima and maxima at which we evaluate the local Majorana polarization shown in Fig. \ref{Graphs}}
	\label{PM_cut}
\end{figure}
\begin{figure}
	\includegraphics[scale=0.25]{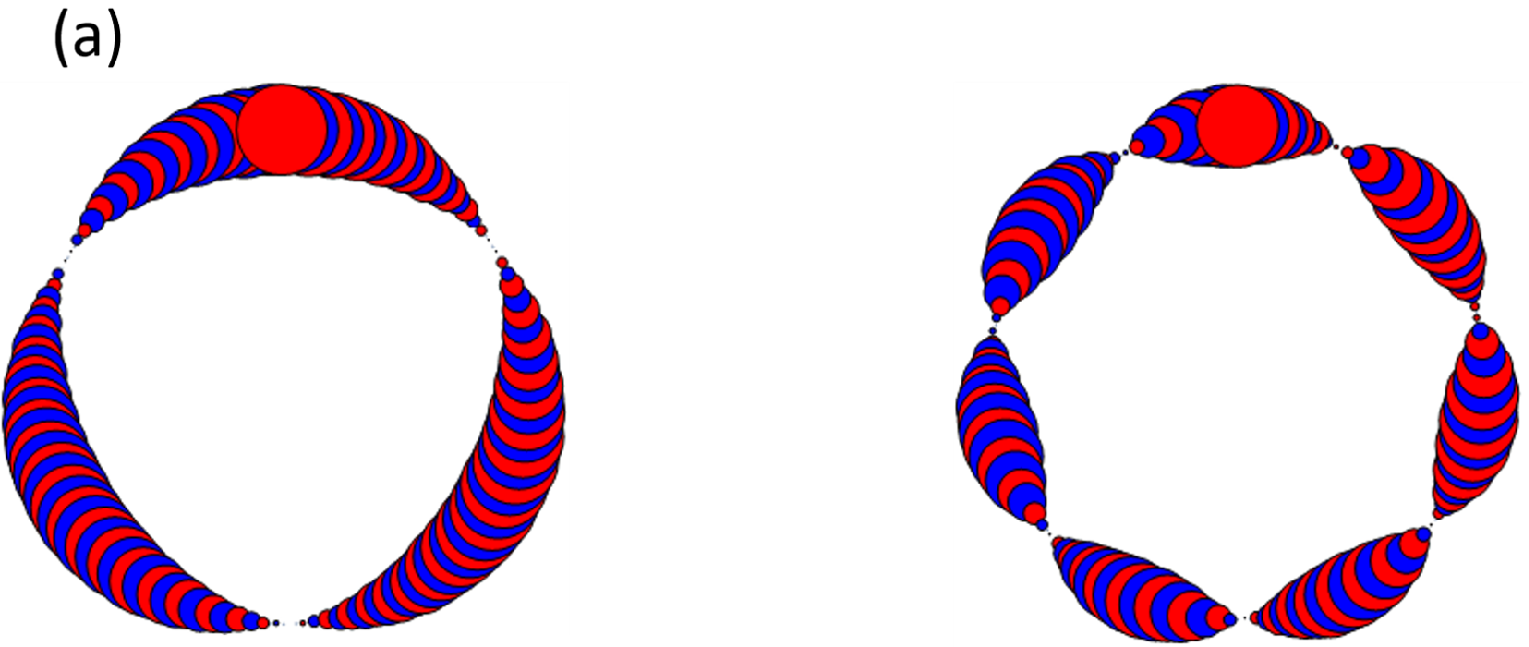}
	\includegraphics[scale=0.25]{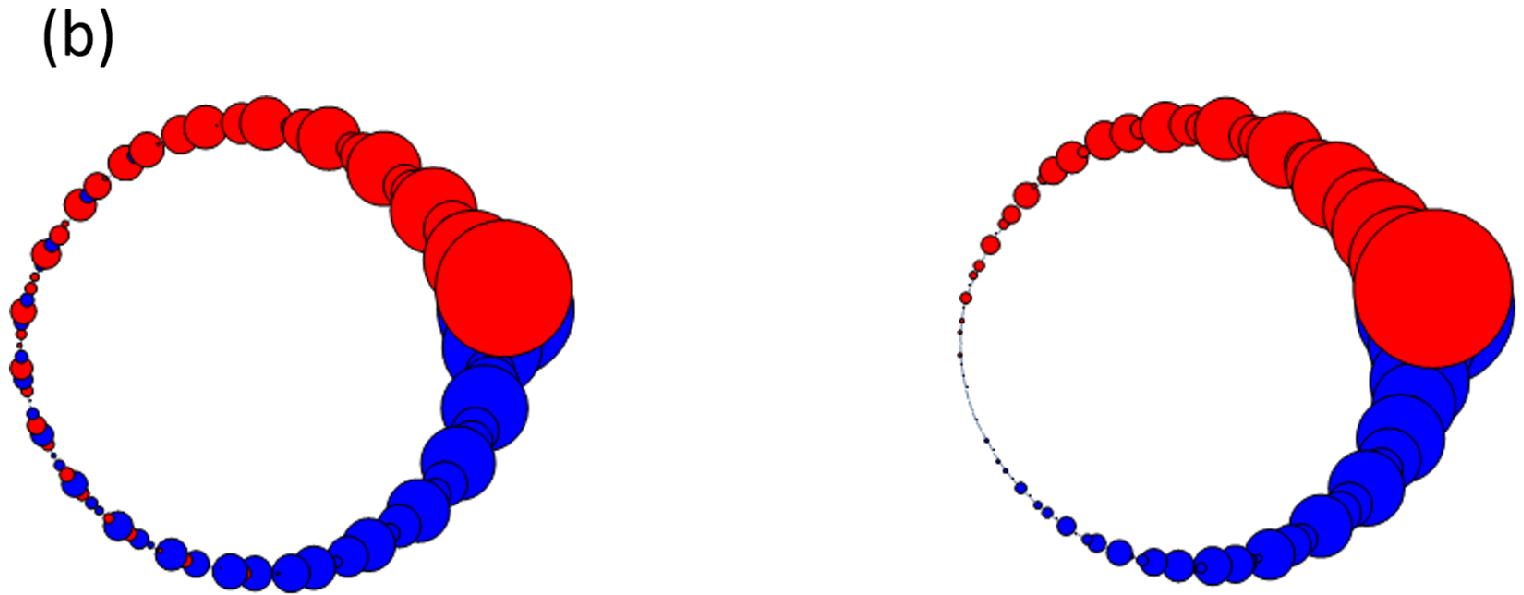}
	\includegraphics[scale=0.25]{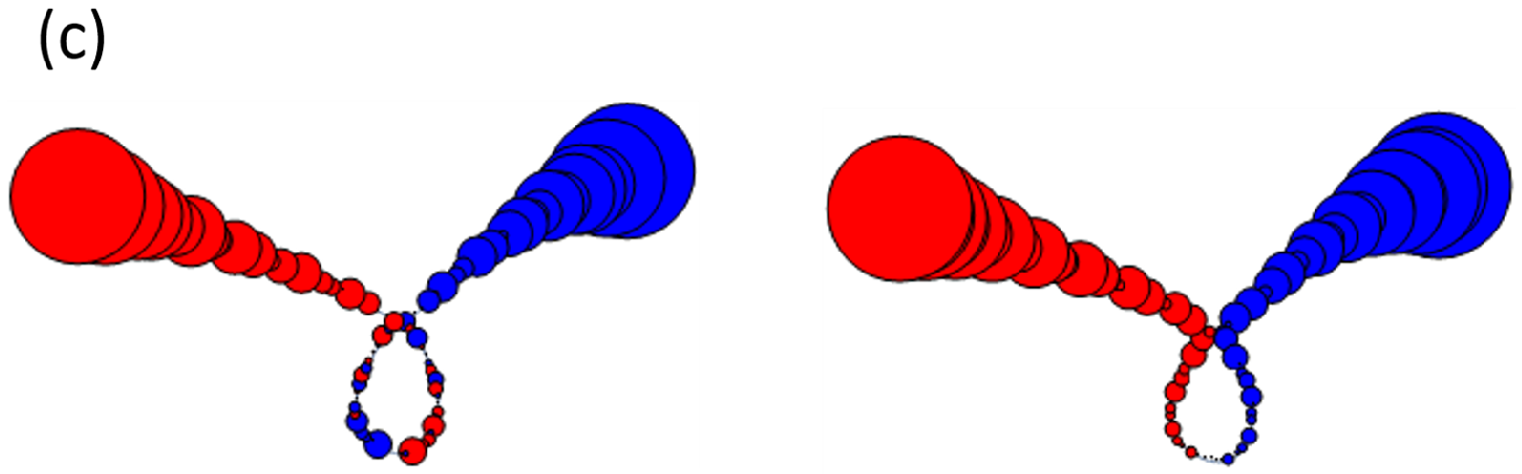}
	\includegraphics[scale=0.25]{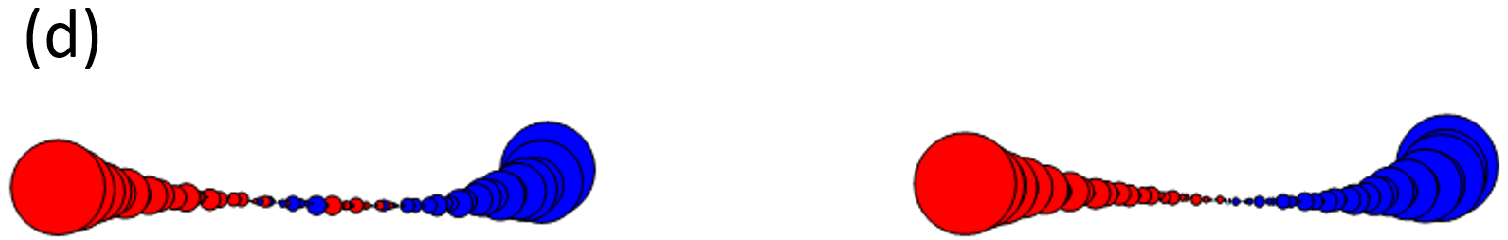}
	\caption{Real space Majorana polarization for the system parameters corresponding to the numbered black circles ($1$, $2$) of Fig. \ref{PM_cut}. Panels (a), (b), (c) and (d) follow the same order of panels of Fig. \ref{PM_cut}. The model parameters have been fixed as: $t = t_d = 1$, $\Delta= 0.02$ and $L = 121$.}
	\label{Graphs}
\end{figure}

\subsection{\label{sec3.2}Majorana polarization}

Another quantity that permits to evaluate the topological phase diagram is the Majorana polarization (MP)  \cite{PhysRevLett.108.096802, PhysRevLett.110.087001, BENA2017349, PhysRevLett.108.096802}. This is a topological order parameter, analogous to the local density of states (LDOS), which measures the quasiparticles weight in the Nambu space. \\
Let us introduce the Nambu representation $\Psi=(c_1,c_1^\dagger,...,c_L,c_L^\dagger)^T$. Accordingly  the Hamiltonian in Eq. (\ref{Hamiltonian}) can be written in the Bogoliubov-de-Gennes form:
\begin{equation}
H=\frac{1}{2}\Psi^\dagger H_{BdG}\Psi
\label{BdGForm}
\end{equation}
where $H_{BdG}$ is a $2 L\times 2L$ matrix being $L$ the number of lattice sites. The eigenstates of $H_{BdG}$ are expressed in the electron-hole basis as $\psi^T=(e_1,h_1,...,e_L,h_L)$ and the local Majorana polarization is defined as:
\begin{equation}
P_M(n)=\int_{-\infty}^\infty P_M(\omega,n) d\omega
\label{Polarization}
\end{equation}
where
$$P_M(\omega,n)=2\sum_{m}\delta(\omega-\epsilon_m)e^{(m)*}_nh_n^{(m)}$$
is the density of MP and $e_n^{m}$ ($h_n^{m}$) refers to the $m$-th eigenstate, while $n$ labels the site. If a state $\psi$ belongs to the particle or hole sector, i.e. $e_n=0$ or $h_n=0$ $\forall$ $n$, the  $P_M$ is indeed zero. On the other hand, the Majorana polarization $P_M=\sum_{n=1}^{L/2}P_M(n)$ of a genuine Majorana state is $\pm 1$. We also note that the system has to satisfy the constraint: $P_M^{tot}=\sum_{n=1}^{L}P_M(n)=0$, because free Majorana monopole cannot exist. In panel (b) of Fig. \ref{PD} we show the topological phase diagram obtained by evaluating the Majorana polarization of the legged-ring system ($t_d\neq 0$) measured in units of the Majorana polarization of the Kitaev chain ($t_d=0$) with the same system length  $L=121$.  We recover qualitatively the same phase diagram of panel (a) with alternating trivial/non-trivial phases. The effect of frustration is also analyzed in Fig. \ref{PM_cut} where we show the MP as a function of the chemical potential by varying the range of the extra hopping: $d=1$ (Kitaev ring), $3$, $40$, $59$ (perturbed Kitaev chain). The case of a Kitaev chain of $121$ sites (red curves) is also plotted for comparison. Going from the Kitaev ring limit (panel (a)) to the perturbed Kitaev chain limit (panel(d)) the MP mean value  increases favoring the non-trivial regime. On the other hand, the alternation of local minima and maxima keeps track of the geometric frustration of the system induced by the long-range hopping. The phenomenology of the frustration is clear when looking at Fig. \ref{Graphs} where the real space Majorana polarization is plotted
in correspondence of the minima and maxima of Fig. \ref{PM_cut} (indicated by the black circles). The size of the circles is proportional to the absolute values of the local MP, while blue and red colors refer to positive and negative values of the MP, respectively.
As shown, the local minima correspond to hybridized Majorana states and the hybridization becomes stronger when $d$ is lower as shown in Fig.4. This is clearly seen in the extreme case of a Kitaev ring (panel (a)) where the polarization is uniformly distributed throughout the system. On the other hand, local maxima correspond to Majorana modes localized at the edges of the legs. Up to now, we have considered the $t_d=t$ case. When the case $t_d \neq t$ is considered, a phase diagram similar to the homogeneous case is obtained (see the Appendix \ref{sec:hopping}.)\\

\section{\label{sec4} Building of a topological frustrated translational invariant system}

\begin{figure}
	\includegraphics[scale=0.3]{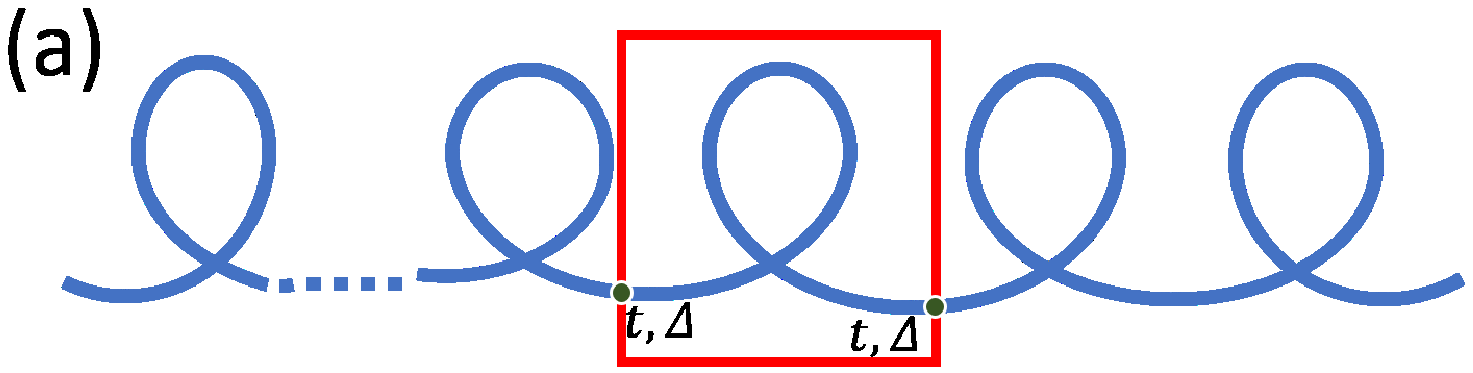}
		\includegraphics[scale=0.33]{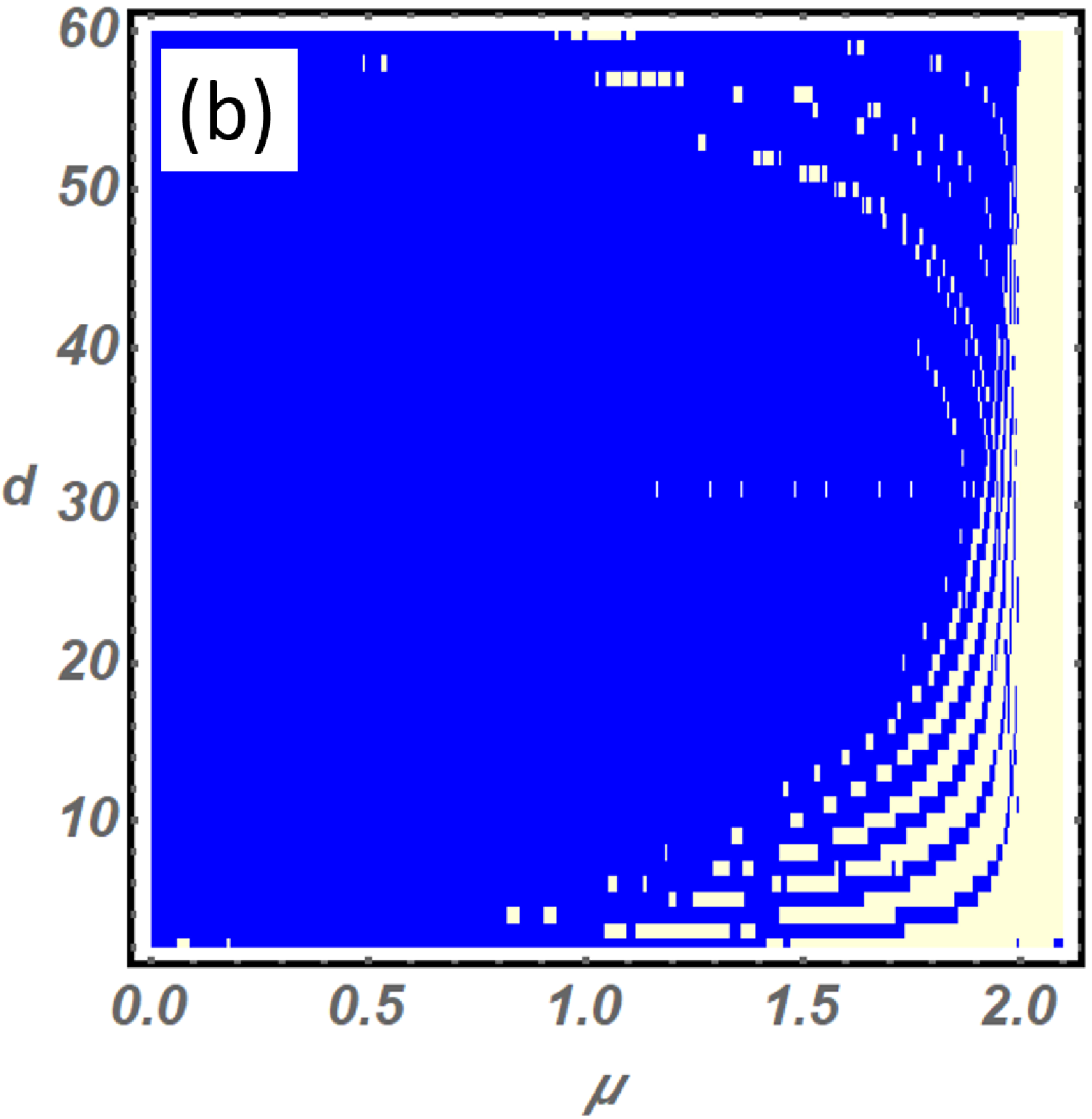}
		\caption{Panel (a): The multiple-tie system with $N$ unit cell. The red square is the n-th unit cell. Panel (b): topological phase diagram in $\mu-d$ plane of the model given by the Majorana number $Q_M$. The topological (trivial) phases correspond to the blue (white) regions. The parameters have been fixed as: $L=121$, $\Delta=0.02$, $t=1$.}
		\label{MKSystem}
\end{figure}

We now consider a multiple-tie system which is the simplest model in which translational invariance coexists with the geometric frustration of the single unit cell.   Beyond the theoretical interest, such a model can describe the multiple loops geometry made by nanotubes \cite{doi:10.1021/nl901260b,PhysRevLett.98.246803}. Thus we define a multiple-tie system with $N$ unit cells each of which having a tie of fixed size $L$ (see Fig. \ref{MKSystem} panel (a)). In the thermodynamic limit $N \to \infty$, translational invariance  is recovered. Imposing periodic boundary conditions, $c_{j,N+1}^\dagger=c_{j,1}^\dagger$, and performing the Fourier transform of the fermionic operators:
 $$c_{j,n}^\dagger=\frac{1}{\sqrt{N}} \sum_k c_{j,k}^\dagger e^{-ikn}$$
 where $k\in [-\pi,\pi]$ is the wave vector, $j$ is the lattice site, while $n$ labels the unit cell $n=1,\dots, N$, the multiple-tie  Hamiltonian can be written in the momentum space as:
 \begin{eqnarray}
 H_{MK}(k)=H_{c}(k)+H_{ic}(k),
 \label{HMK}
 \end{eqnarray}
where $H_c$:
\begin{eqnarray}
H_c(k)=&&\frac{1}{2}\sum_{j}\sum_{k} \big[-\mu(c_{j,k}^{\dagger} c_{j,k}+c_{j,-k}^{\dagger} c_{j,-k})+\nonumber\\
&&-t(c_{j,k}^{\dagger} c_{j+1,k}+c_{j,-k}^{\dagger} c_{j+1,-k}+h.c.)+\nonumber\\
&&\Delta(c_{j+1,k}^{\dagger} c^{\dagger}_{j,-k}+c_{j+1,-k}^{\dagger} c_{j,k}^{\dagger}+h.c.)\big]
\label{Hunitcell}
\end{eqnarray}
while $H_{ic}$ is given by:
 \begin{eqnarray}
&& H_{ic}(k)=\frac{1}{2}\sum_{k} \big[-t(c_{1,k}^{\dagger} c_{L,k}e^{-ik}+c_{1,-k}^{\dagger} c_{L,-k}e^{ik}+ \nonumber\\
&& +h.c.)+\Delta(c_{1,k}^{\dagger} c^{\dagger}_{L,-k} e^{-ik}+c_{1,-k}^{\dagger} c_{L,k}^{\dagger} e^{ik}+h.c.)\big].
 \label{Hcoupling}
 \end{eqnarray}
Since the Hamiltonian is now translational invariant, one can compute the topological bulk invariant corresponding to the Majorana number introduced by Kitaev \cite{Kitaev_2001}. Let us first introduce the Majorana operators in $k$-space: $a_{j,k}=c_{j,k}+c_{j,-k}^\dagger,\ \ b_{j,k}=(c_{j,k}-c_{j,-k}^\dagger)/i$ in terms of which the Hamiltonian becomes:
 \begin{eqnarray}
 H_{MK}=\frac{i}{2}\Psi^\dagger_M\left[ H_T+(T_1 e^{ik}+D_1 e^{-ik}+h.c.)\right] \Psi_M\nonumber
 \end{eqnarray}
where $\Psi_M=(a_{1,-k},b_{1,-k},\dots,a_{L,-k},b_{L,-k})^T$ and $a^\dagger_{j,k}=a_{j,-k}$, $b^\dagger_{j,k}=b_{j,-k}$. In the new basis, $H_T$, $T_1$ and $D_1$ are  $2L \times 2L$ matrices whose structure is defined by Eq. (\ref{Hunitcell}) and Eq. (\ref{Hcoupling}). More specifically:
\[T_1=\begin{pmatrix}
&&&&&&&.\\
&&&&&&.\\
&&&&&.\\
&&&.\\
0&-t\\
t&0\\
\end{pmatrix},\ \ D_1=\begin{pmatrix}
&&&&&&&.\\
&&&&&&.\\
&&&&&.\\
&&&.\\
0&\Delta\\
\Delta&0\\
\end{pmatrix},\]
where the dots stand for null elements. The Majorana number, is defined as:  $Q_M=Sign\left[Pf M(0) \right] Sign\left[Pf M(\pi) \right]$, where $Pf M$ is the Pfaffian of the Hamiltonian $M(k)=H_T+(T_1 e^{ik}+D_1 e^{-ik}+h.c.)$ evaluated at the points $k=0$, $\pi$ in the momentum space. The computation proceeds as follows. We first reduce the Hamiltonian to a canonical form: $M'=U M U^T$, by means of an orthogonal $2L \times 2L$ matrix $U$ whose rows are the eigenvectors of $M$:
\[M'=\begin{pmatrix}
0&\lambda_1&&&\\
-\lambda_1&0\\
&&&.\\
&&&&.\\
&&&&&.\\
&&&&&&0&\lambda_{2L}\\
&&&&&&-\lambda_{2L}&0\\
\end{pmatrix},\\ PfM'=\lambda_1 \dots \lambda_{2L}\]\\
then using the Pfaffian property: $Pf(UMU^T)=det[U]Pf(M)$, the Majorana number can be recast in the form:
\begin{equation}
Q_M=Sign\left[\frac{PfM'(0)}{det(U)} \right] Sign\left[\frac{PfM'(\pi)}{det(U)} \right],
\end{equation}
which is evaluated numerically.
Panel (b) of Fig. \ref{MKSystem} shows the phase diagram in $d-\mu$ plane of a multiple-tie system when the single unit cell has size $L=121$. The topological phases correspond to $Q_M=-1$ (blue regions) while the trivial phases correspond to $Q_M=1$ (white regions). We note that the  trivial/non-trivial phases sequence is still present but only for values of the chemical potential close to the value $\mu=2t$ where the topological phase transition is expected for a Kitaev chain. The presence of trivial phases close before $\mu=2t$ is essentially due to the frustration of the single unit cell. Bulk-edge correspondence is explicitely proven for a sytem of reduced size in the Appendix \ref{sec:example}.

\section{\label{sec5}Conclusions}

In conclusion, we have presented an analysis of the topological phase diagram of a Kitaev chain with geometric frustration caused by the presence of a long-range hopping (Kitaev tie). Due to the breaking of the translational invariance, the bulk-edge theorem cannot be used. Thus we have resorted to a real space method based on a generalization of the transfer matrix method. By the calculation of the transfer matrix we have studied the emergence of localized Majorana wave functions at the edge of the legs.
We have found that the geometric frustration gives rise to an interstitial-like behavior of the topological phase diagram in which non-trivial phases alternate with trivial ones at varying the chemical potential and the range of the extra hopping, controlled by the parameter $d$. We have also shown that the non-trivial phases enlarge and become dominant when the perturbed Kitaev chain limit (i.e. large values of parameter $d$) is considered. The same interstitial-like character of the topological phase diagram emerges when the Majorana polarization is considered. Moreover, we have considered a multiple-tie system in which  translational invariance coexists with frustration effects. In the latter case, the effect of geometric frustration is  reduced and the bulk-edge correspondence has been proven. \\
The effect of geometric frustration studied in this work has been poorly investigated in connection with topological phase transitions. Despite this, topological frustration could be a relevant ingredient to design proof-of principle nanodevices. In this respect,  looped or flexible nanowires, such as e.g. carbon nanotubes, are the main testbed to prove the topological frustration physics described here.

\begin{acknowledgments}
R.C. acknowledges the Project QUANTOX (QUANtum Technologies with 2D-OXides) of QuantERA ERA-NET Cofund in Quantum Technologies (Grant Agreement N. 731473).
\end{acknowledgments}
\appendix

\section{Effect of the hopping strength}
\label{sec:hopping}

\begin{figure}
	\includegraphics[scale=0.35]{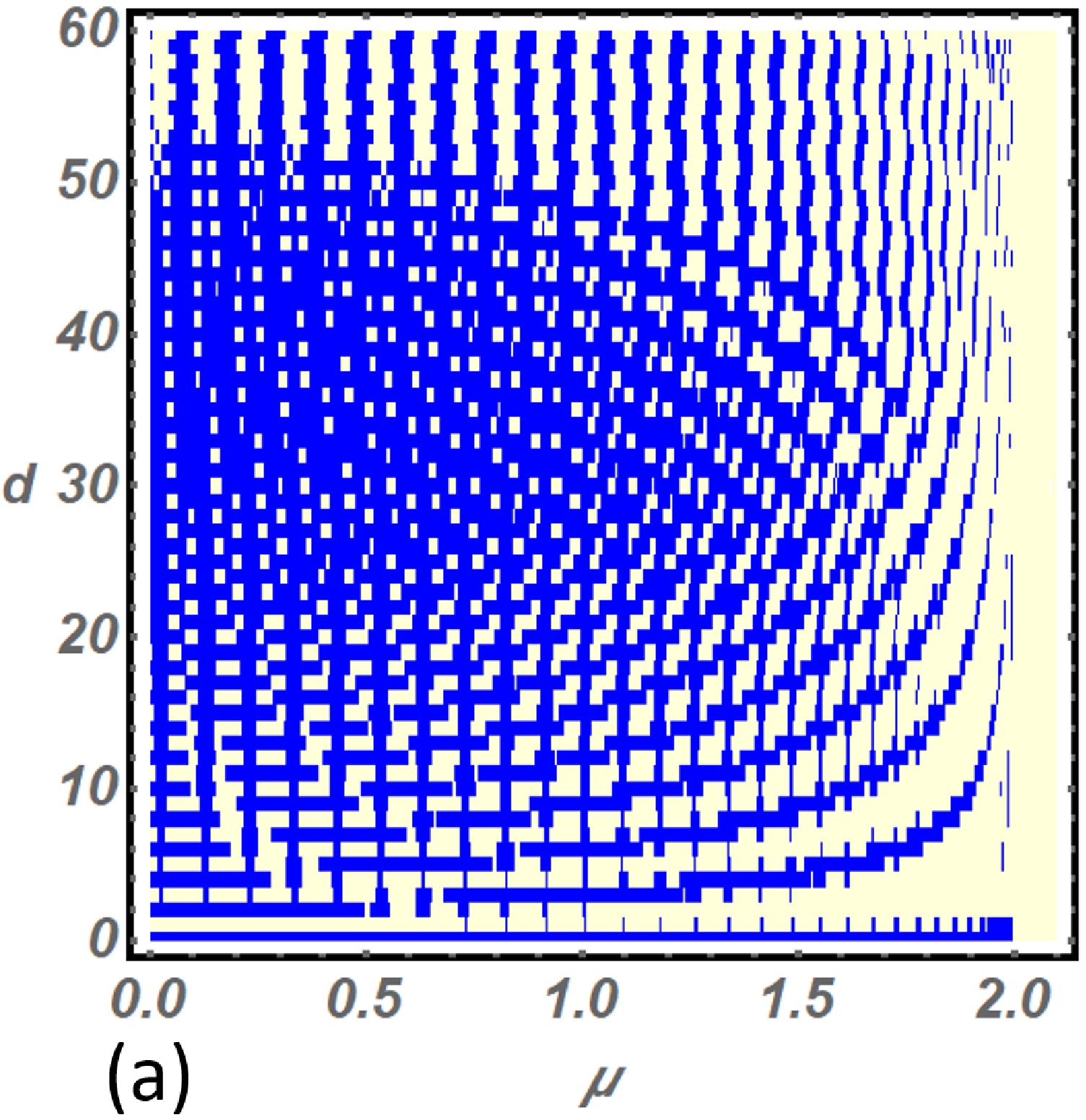}
	\includegraphics[scale=0.45]{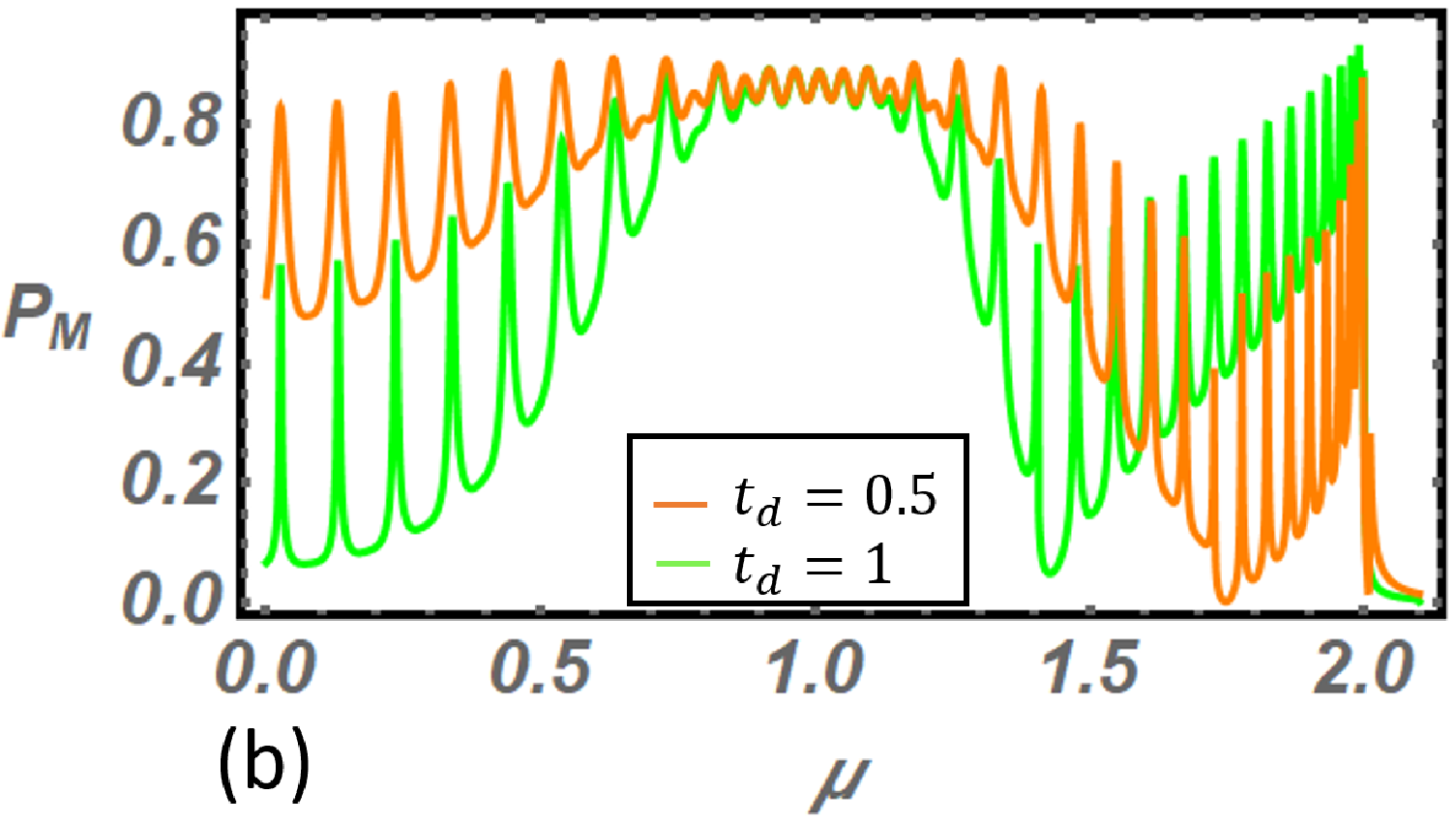}
	\caption{(a) Topological phase diagram in the $\mu-d$ plane for $t/t_d=0.5$. (b) Majorana polarization as a function of chemical potential $\mu$ for $t_d=1$ (green curve) and $t_d=0.5$ (orange curve). The other parameters are: $t=1$, $\Delta=0.02$, $d=3$, $L=121$.}
	\label{inhomogeneousCase}
\end{figure}

In the main text, we have shown the effect of geometric frustration by varying the range of the extra hopping. Here we investigate the effect of changing the amplitude of  long range hopping so that  $t_d\neq t$.  This analysis is performed in Fig. \ref{inhomogeneousCase}. Since the transfer matrix and MP methods provide compatible results, we restrict our analysis to the Majorana polarization. Panel (a) of Fig.  \ref{inhomogeneousCase} shows that for $t_d/t=0.5$ the extension of non-trivial phases is increased compared to the homogeneous case ($t_d=t=1$) reported in  Fig. \ref{PD} (panel (b)). In particular, panel (b) of Fig. \ref{inhomogeneousCase} shows the Majorana polarization as a function of chemical potential for $t_d=1$ and $t_d=0.5$ when $d=3$, i.e. a quasi-ring with short legs. At decreasing the strength of the hopping between sites $\alpha$ and $\beta$ the MP mean value increases, favoring the non-trivial regime.

\section{Bulk-edge correspondence for a multiple-tie system}
\label{sec:example}
\begin{figure}
\includegraphics[scale=0.35]{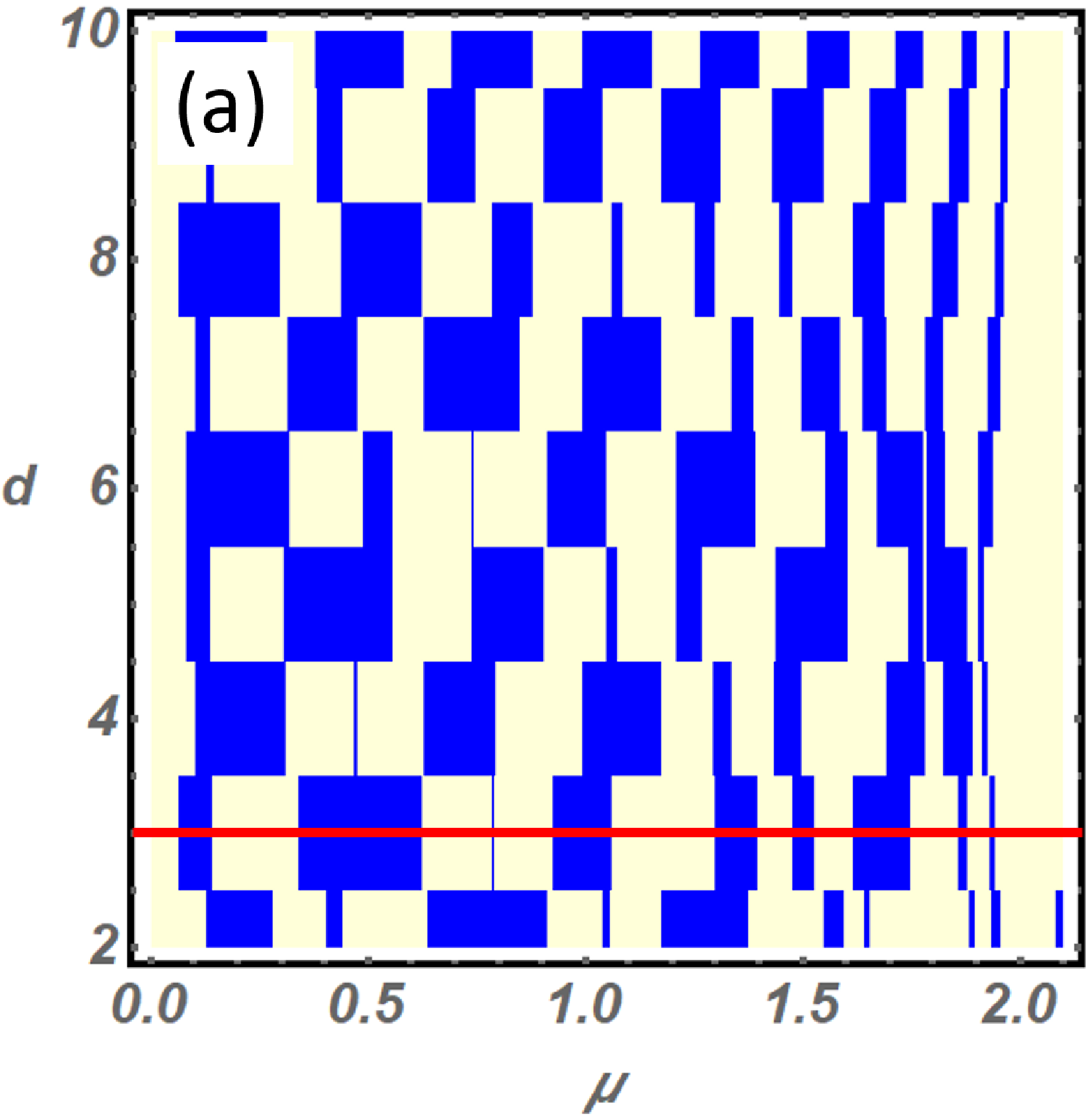}
\includegraphics[scale=0.45]{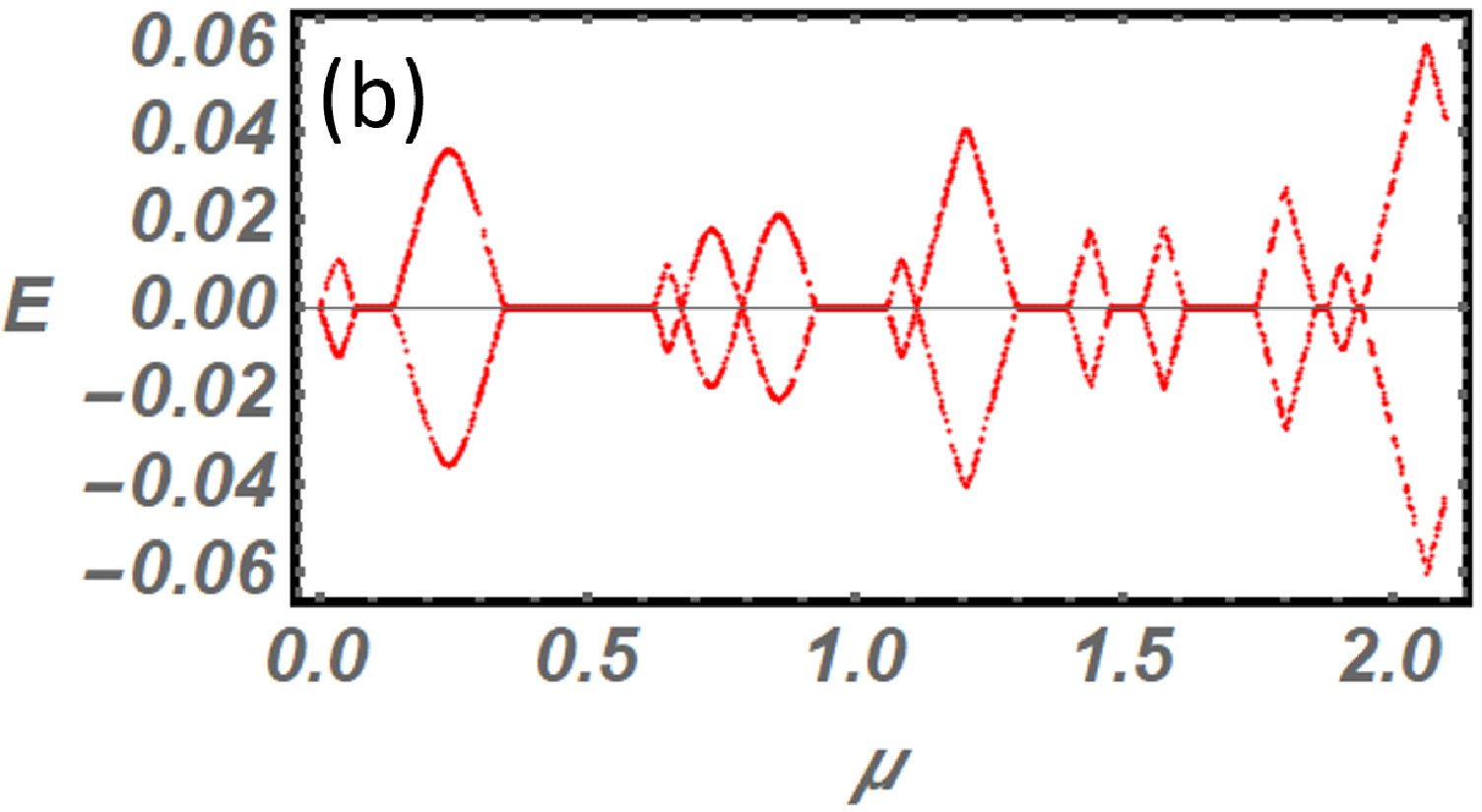}
\caption{(a): Topological phase diagram in $\mu-d$ plane of a multiple-tie system of $30$ unit cells and $20$ sites per unit cell obtained by the Majorana number $Q$. The topological (trivial) phases correspond to the blue (white) regions. (b): energy eigenvalues of the system as a function of chemical potential $\mu$ corresponding to the horizontal red cut of panel (a). The other parameters have been fixed as: $\Delta=0.02$, $t=1$.}
\label{l=20}
\end{figure}

\begin{figure}[h]
\includegraphics[scale=0.35]{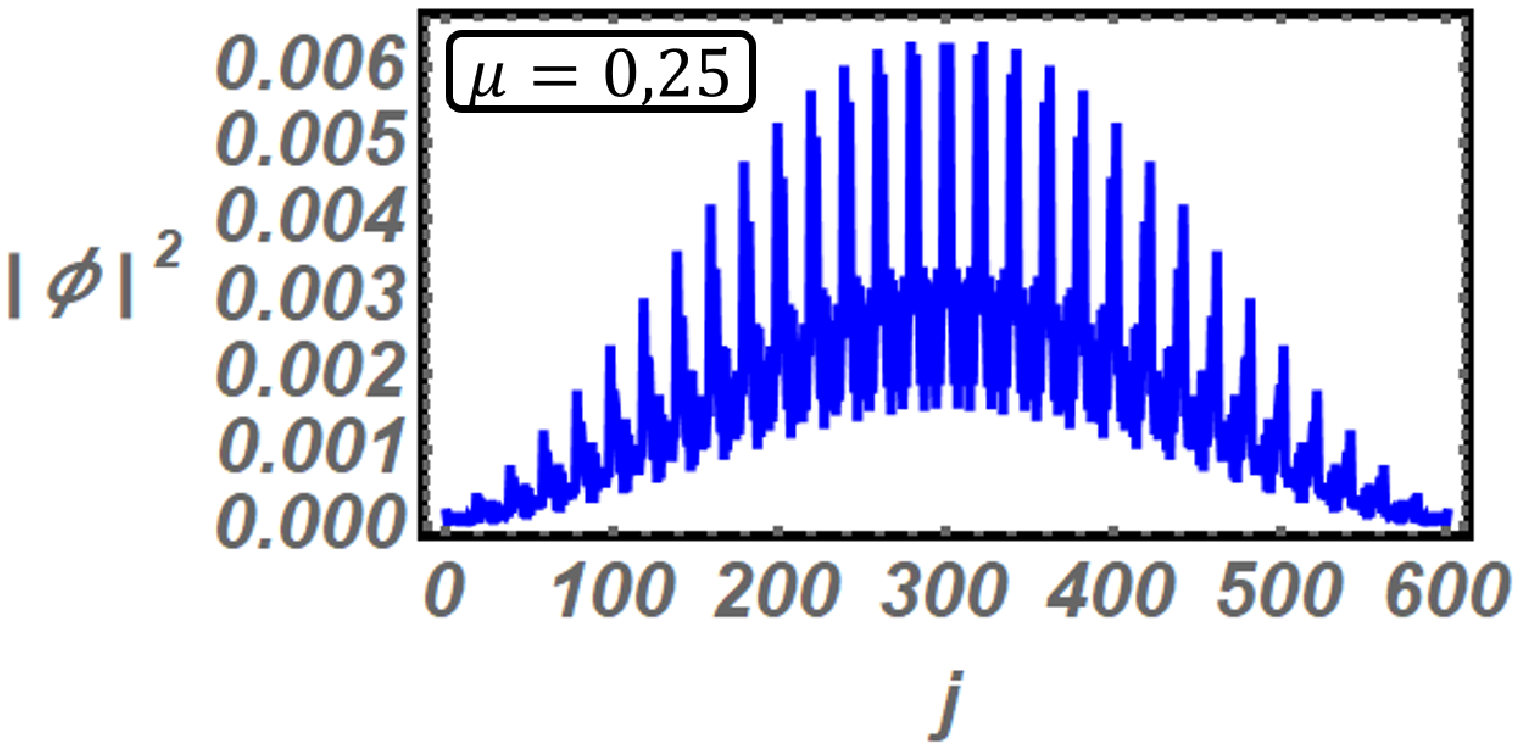}
\includegraphics[scale=0.35]{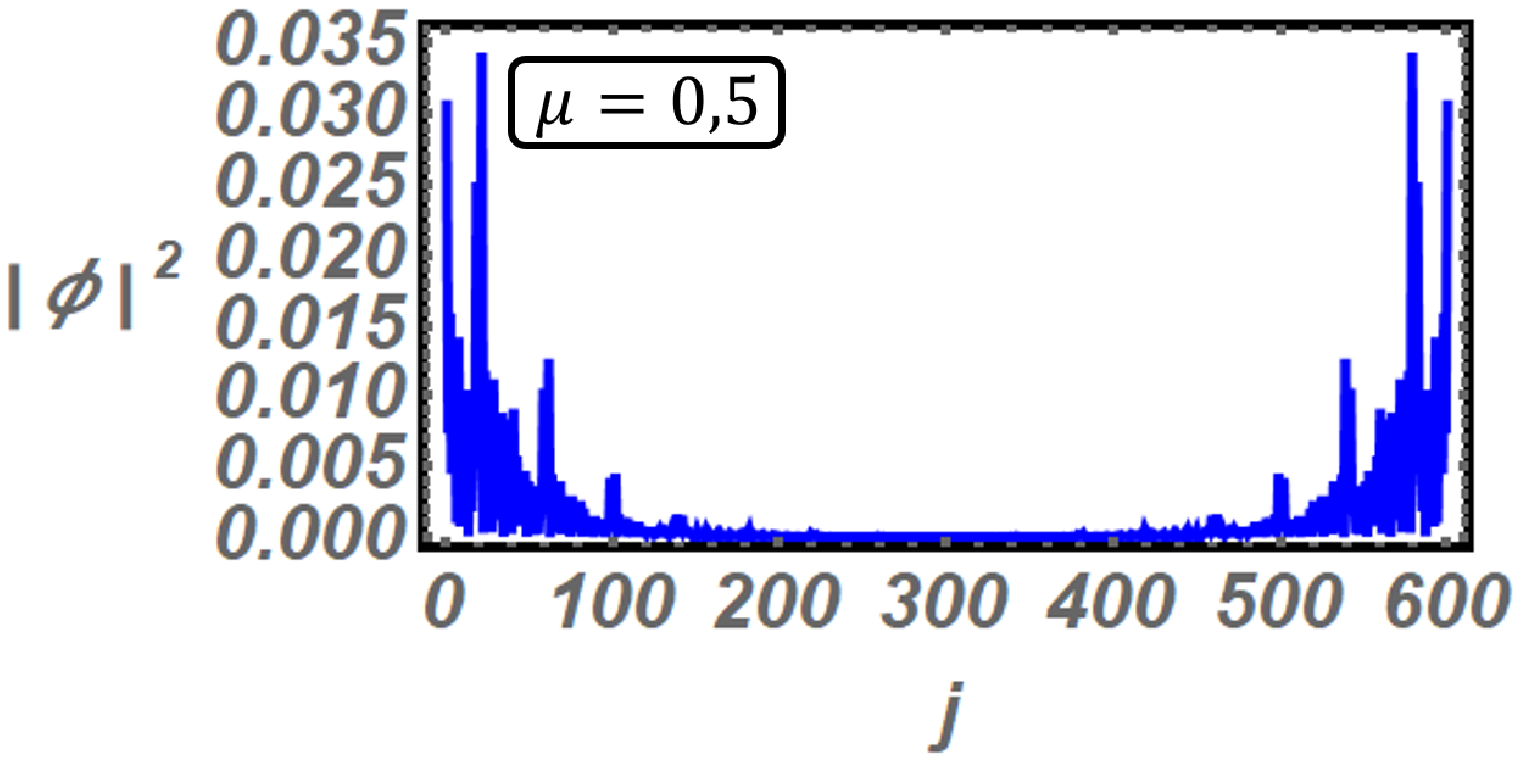}
\includegraphics[scale=0.35]{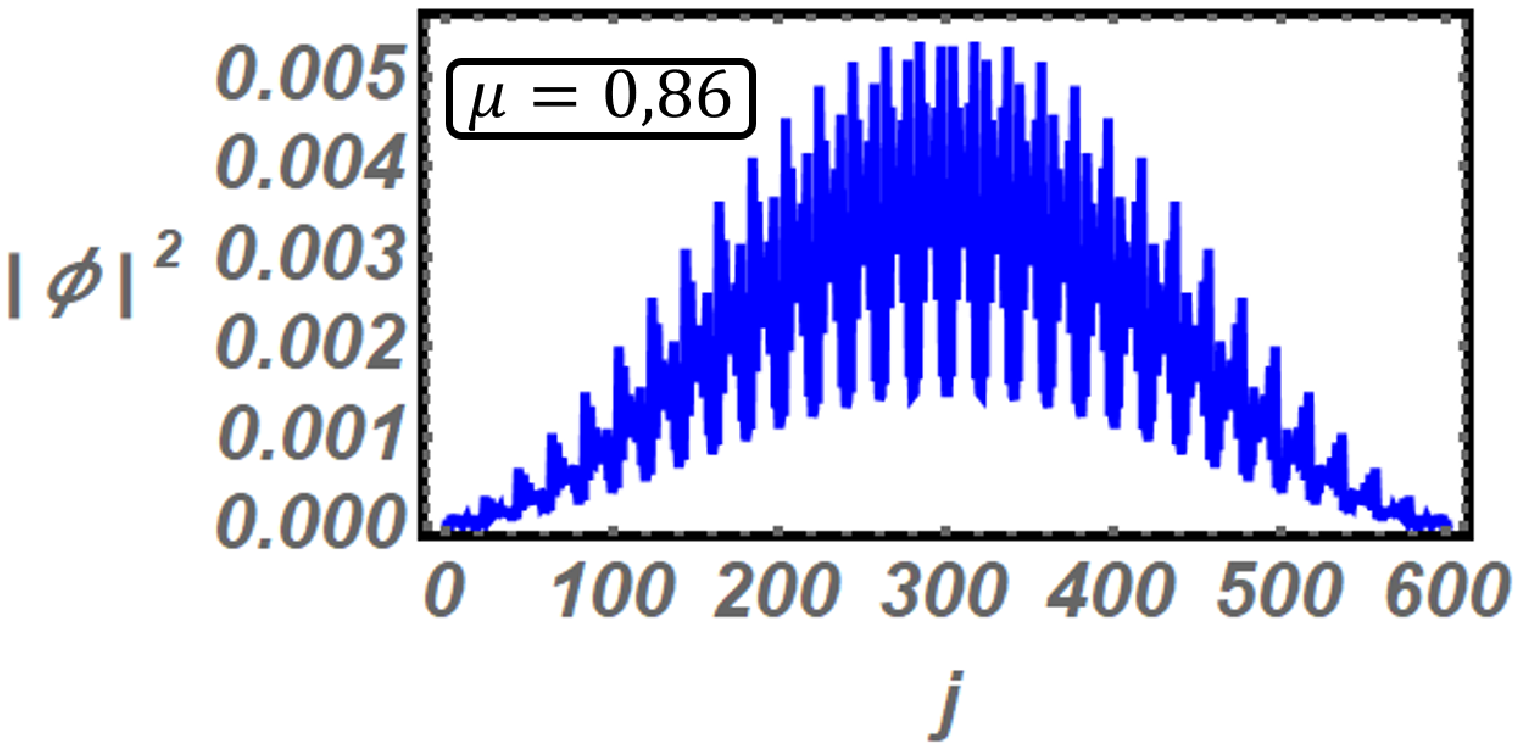}
\includegraphics[scale=0.35]{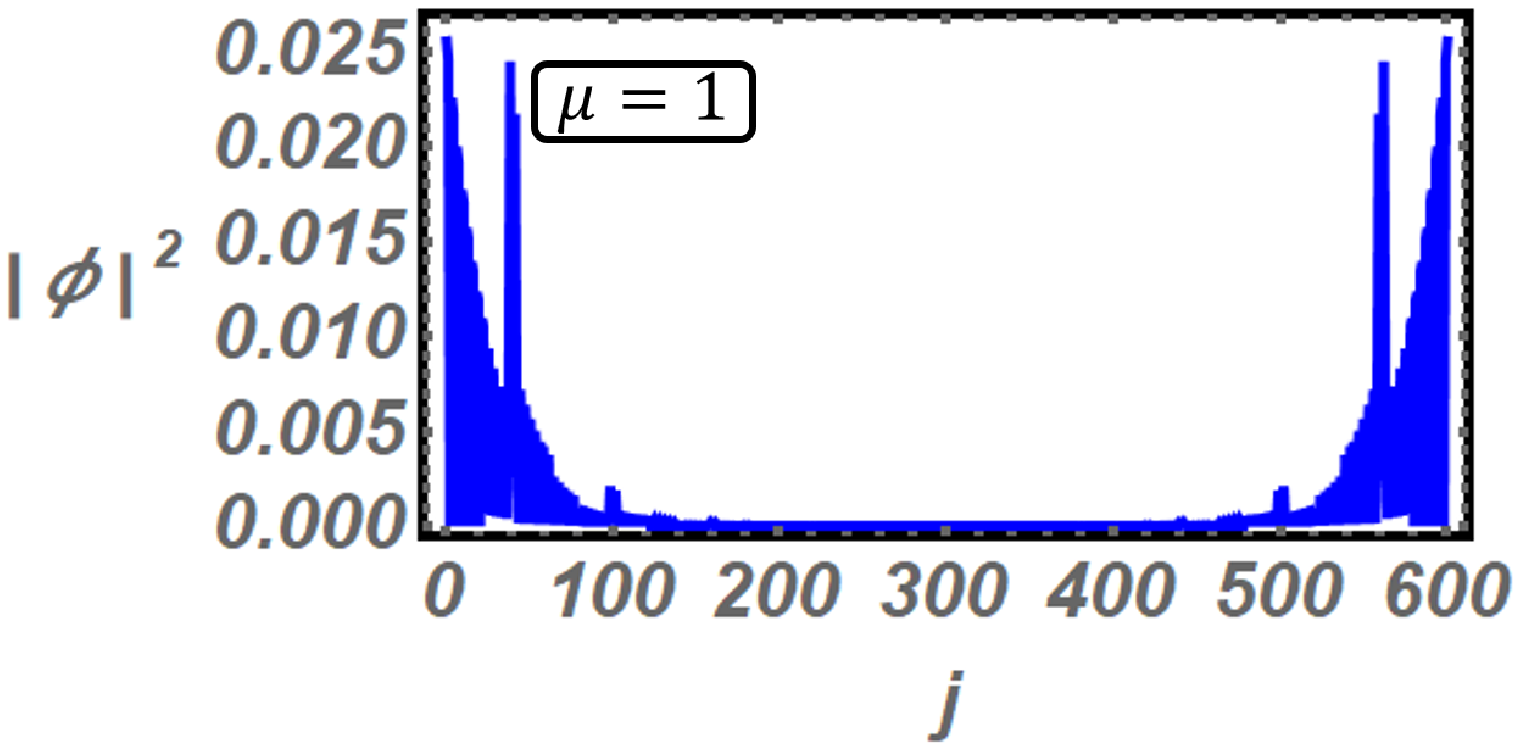}
\caption{Amplitude of the lowest energy modes of the multiple-tie system with $N=30$ unit cells and with $L=20$ sites per unit cell, as a function of position $j$ and four different values of the chemical potential $\mu$ along the red line of Fig.\ref{l=20}(a). The other parameters have been fixed as: $\Delta=0.02$, $t=1$.}
\label{Sqmodulus}
\end{figure}

In this Appendix we show bulk-edge correspondence for a multiple-tie system made of $30$ unit cells. In particular, in Fig. \ref{l=20} (a), we show  the phase diagram of the translational invariant multiple-tie system having $20$ sites per unit-cell. The phase diagram has been obtained by exploiting the band topological invariant. It shows a checkerboard pattern which is reminiscent of the topological frustration of the single unit cell. Panel (b) of Fig. \ref{l=20} shows the lowest energy eigenvalues corresponding to the red horizontal cut of panel (a). The correspondence between trivial and non-trivial phases  is clearly visible in Fig. \ref{l=20} (b).
In order to get further insight, in Fig. \ref{Sqmodulus}, we show the localization properties of the wavefunction for a trivial/topological phase sequence moving the chemical potential along the red line of Fig.\ref{l=20}(a). This analysis directly shows that localized states correspond to the gapless points in Fig. \ref{l=20}(b), while trivial states correspond to the gapped ones.

\newpage
\newpage

\bibliographystyle{apsrev4-1}

\begin{thebibliography}{35}
\bibitem [{not()}]{note}%
  \BibitemOpen
  \href@noop {} {}\bibinfo {note} {Let us note that for finite-sized systems
  the ground state degeneracy is broken by an exponentially small splitting due
  to a weak interaction between Majorana modes described by an effective
  Hamiltonian: $H_{eff}=\frac{i}{2}J a_1 b_N$ with $J \propto
  e^{-\frac{L}{l_0}}$ thus causing a lifting of the energy from zero. This
  implies that $T_{11}\approx0$ for very weakly interacting Majorana
  fermions.}\BibitemShut {Stop}%
\end{thebibliography}
%

\end{document}